\documentclass[traditabstract,a4]{aa}  

\usepackage{epsfig}
\usepackage{graphicx}
\usepackage{epstopdf}
\usepackage{color}  
\usepackage{array}   
\usepackage{units}  
\usepackage[breaklinks,colorlinks, citecolor=blue]{hyperref}
\usepackage{natbib}  
\usepackage{multirow}   
\usepackage{amsmath,amssymb}  
\usepackage[english]{babel}
\usepackage[latin1]{inputenc}
\usepackage[T1]{fontenc}
\usepackage{longtable,lscape}
\usepackage{footnote}

\usepackage{txfonts}
\usepackage[toc,page]{appendix} 

\begin{document}


\title{Extraction of $^{12}$CO and $^{13}$CO maps from {\it Planck} data}
\author{G.~Hurier}

\institute{
  Centro de Estudios de F\'isica del Cosmos de Arag\'on (CEFCA), Plaza de San Juan, 1, planta 2, E-44001, Teruel, Spain 
  \label{CEFCA} \\
  \email{hurier.guillaume@gmail.com}
}

   \abstract{Rotational transition lines of CO are one of the major tracers used to study star forming regions and Galactic structures.
   A large number of observations of CO rotational lines are covering the galactic plane, recently the {\it Planck} collaboration release the first full sky coverage of J=1-0, J=2-1, and J=3-2 CO lines at a resolution of 10', 5', and 5' FWHM. However, the measured signal in {\it Planck} detectors is integrated over large bandpass $(\Delta \nu)/\nu \simeq 0.2$. Consequently, the derived CO products are composite maps including several rotational lines. In the 100 GHz {\it Planck} channel, the two main lines are the J=1-0 transitions of $^{12}$CO and $^{13}$CO at 115 and 110 GHz respectively.
   In the present paper, we present and applied a method to construct separate CO integrated intensity maps for the two isotopes.
   The measurement of the $^{13}$CO rotational transitions provide an unprecedented all-sky view of the Galaxy for this isotope.} 


\authorrunning{G.Hurier et al.}
\titlerunning{$^{12}$CO and $^{13}$CO maps from {\it Planck} data}

\maketitle

\section{Introduction}

This paper describes the construction and validation of full-sky carbon monoxyde (CO) maps for $^{12}$CO and $^{13}$CO isotopologues from {\it Planck} data.\\
The interstellar medium (ISM) constitutes about 10-15\% of
the total mass of the Milky Way. The ISM is a composed by atomic
and molecular gas, both contain around 50\% of its total mass \citep[see][for a review]{fer01,cox05}. The
cold molecular gas, that represent a small fraction of it, is essentially located close to the Galactic plane. 
Molecular clouds are hosting star formation, and as such, play a major role in the interstellar matter cycle.\\
Molecular clouds were discovered via the rotational
emission line J=1-0 of carbon monoxide in its fundamental
electronic and vibrational levels \citep{wil70,pen72}. Contrary to the atomic component of the neutral
ISM, which is directly observable via the spin-flip HI, 21 cm
line, the bulk of molecular hydrogen is not directly observable
in molecular clouds. CO is abundant, easily excited by
collisions with H$_2$, and easily observable from the ground, it is
therefore considered as a good tracer of the molecular component of the ISM.\\

Thanks to bandpass differences between {\it Planck} bolometers in a given frequency channel, large-scale surveys of J=1-0, J=2-1, and J=3-2 CO lines at a resolution of 10', 5', and 5' FWHM angular resolution have been obtained from the {\it Planck} satellite data \citep{PlanckCO}. However, these CO maps contain several lines, mainly from $^{12}$CO, but also from $^{13}$CO, and other lines for the J=3-2 CO map extracted from {\it Planck} 353 GHz channel.\\
Extracting CO maps from {\it Planck} data is of interest for handling systematic effect in the map-making of polarized map. 
The bandpass differences that induce significant variations of the CO transmission in {\it Planck} polarized detectors induce confusion between intensity and polarization in the map-making process. 
Considering that $^{12}$CO and $^{13}$CO are emitting lines at different frequencies, their transmissions in {\it Planck} detectors is different. 
Therefore, to properly account for the CO leakage into {\it Planck} polarized map both $^{12}$CO and $^{13}$CO have to be considered independently.\\
Large-scale surveys of the J=1-0 line of $^{12}$CO, but also of $^{13}$CO, have been carried out with meter-sized radio telescopes. 
\citet{dam01} conducted the most complete $^{12}$CO(1-0) survey, which covers the Milky Way at Galactic latitudes $\| b \| \leq 30^{o}$, with an effective spatial resolution of $0.5^{\rm o}$.
In addition, there exist a wealth of smaller $^{12}$CO(1-0) and $^{13}$CO(1-0) line observations \citep[see e.g.,][]{jac06,rid06,nar08}.\\ 
Observations of CO, for the Orion and Monoceros regions, by \citet{mag85,har98,mag00} have revealed the existence of molecular clouds at
Galactic latitudes up to 55$^{o}$. Which implies that the effect of CO on {\it Planck} map-making is also present at high galactic latitudes.
However, these high-latitude observations provide only a limited view of the $\| b \| > 30^o$ sky.\\


The J=1-0 $^{12}$CO and $^{13}$CO rotational transition lines lie within the 100 GHz spectral bands of the HFI instrument. 
In this paper, we extract full-sky $^{12}$CO and $^{13}$CO maps for this transition from {\it PLanck}-HFI data using a component separation
method. The data we used to construct and validate the maps are presented in Sect.~\ref{secdat}. In Sect.~\ref{secmet}, we provide a description of the methodology used to extract the CO maps. 
Comparison of these component separated CO maps with ancillary data is detailed in Sect.~\ref{seccomp}.
Uncertainties and foreground contamination on those maps are discussed in Sect.~\ref{seccar} and future improvements for these maps in Sect.~\ref{secdis}.
Finally, we draw conclusions in Sect.~\ref{seccon}.

\section{Data}
\label{secdat}
\subsection{{\it Planck} data}

This paper is based on the {\it Planck} full mission data release 2\citep{PR2}, it uses the full-sky maps from two detector sets (\emph{detset}) build from the observation of the sky at 100 GHz\footnote{\url{https://irsa.ipac.caltech.edu/data/Planck/release_2/all-sky-maps/maps/HFI_SkyMap_100-ds1_2048_R2.02_full.fits}}\footnote{\url{https://irsa.ipac.caltech.edu/data/Planck/release_2/all-sky-maps/maps/HFI_SkyMap_100-ds2_2048_R2.02_full.fits}}, and it uses the type-1 CO public map\footnote{\url{https://irsa.ipac.caltech.edu/data/Planck/release_2/all-sky-maps/maps/component-maps/foregrounds/HFI_CompMap_CO-Type1_2048_R2.00.fits}} \citep{PlanckCO} derived from {\it Planck} release 2 with the MILCA method \citep{hur13}.
We used the maps in their HEALPix pixelization from \citep{gor05}, with $N_{\rm side} = 2048$ (1.7' arcmin pixels).
These maps are given in $K_{\rm CMB}$ units for the temperature \emph{detset} maps at 100 GHz and in K$_{\rm RJ}$.km/s for the type-1 CO map.
We also made used of half-mission {\it Planck} maps for detector sets. These maps are constructed from subsets of observations and enable the construction of noise maps free from astrophysical emissions.
In the following, we assume circular gaussian beams with a FWHM of 9.88' at 100 GHz to describe the sky signal transfer function.\\

\subsection{Ancillary data}
We also made use of additional CO surveys to calibrate {\it Planck} response to CO emission and to validate the {\it Planck}-data based $^{12}$CO(J=1-0) and $^{13}$CO(J=1-0) maps.\\
In particular, for the calibration of {\it Planck} map responses to CO lines we used
\begin{itemize}
\item[$\bullet$] the $^{12}$CO(J=1-0) by \citet{dam01} that cover the entire galactic plane over a band of 4-10 degrees in latitude,
\item[$\bullet$] the $^{12}$CO(J=1-0) and $^{13}$CO(J=1-0) Boston University FCRAO Galactic Ring Survey \citep[BU FCRAO GRS,][]{jac06} mapping part of the galaxy with a 45" angular resolution,
\end{itemize}
whereas we used
\begin{itemize}
\item[$\bullet$] the FCRAO Ophiuchus and Perseus molecular clouds CO surveys \citep{rid06} at a 45" angular resolution,
\item[$\bullet$] the FCRAO survey of the Taurus molecular cloud \citep{nar08}, covering 96 square degrees at a 45" angular resolution,
\end{itemize}
as reference CO surveys to validate the CO-maps derived from {\it Planck} data combining the MILCA Type-1 CO maps with \emph{detset} maps at 100 GHz.
For all these reference CO surveys we consider the CO integrated intensity. We also used the map from \citet{dam01} to estimate CO Intensity weighted mean velocity on the line of sight which provides the effective frequency at which is seen the signal in the {\it Planck} satellite rest-frame.

\section{Methodology}
\label{secmet}

\subsection{Data model}

Our set of 3 maps (Type-1 CO, \emph{detset}-1, \emph{detset}-2): $T_{\rm CO}$, $T_{\rm DS1}$, and $T_{\rm DS2}$ can be modeled as,
\begin{align}
T_i =  A^{i}_{ ^{12}{\rm CO}} W_{^{12}{\rm CO}} + A^{i}_{ ^{13}{\rm CO}} W_{^{13}{\rm CO}} + A^{i}_{100{\rm GHz}} T_{100{\rm GHz}},
\end{align}
with $A^{i}_{X}$ the transmission coefficients of the component $X$ in the map $i$, $W_{^{12}{\rm CO}}$ and $W_{^{13}{\rm CO}}$ the $^{12}{\rm CO}$ and $^{13}{\rm CO}$ intensity, and $T_{100{\rm GHz}}$ the sky emission at 100 GHz mainly composed by the CMB, thermal dust, Sunyaev-Zel'dovich effect, synchrotron, and free-free emissions. In our situation, these 100 GHz sky components present the same response in all {\it Planck} 100 GHz channel detectors.
Thus, we have a set of three maps that are composed of three varying components.
By construction the transmission $A^{i}_{100{\rm GHz}}$ is equal to unity for \emph{detset} maps and null for the type-1 CO map. 
As a consequence, to remove the 100 GHz sky contribution from the \emph{detset} maps, we only consider their difference. This \emph{detset}-difference map is also a CO map, however it present different weighting over the 100-GHz {\it Planck} detectors, and is therefore presenting a different response to both $^{12}$CO and $^{13}$CO than the official type-1 CO maps.
Finally, we have a set of two maps that are described by two components: $^{12}$CO and $^{13}$CO. 

\begin{figure}[!th]
\begin{center}
\includegraphics[angle=90,scale=0.35]{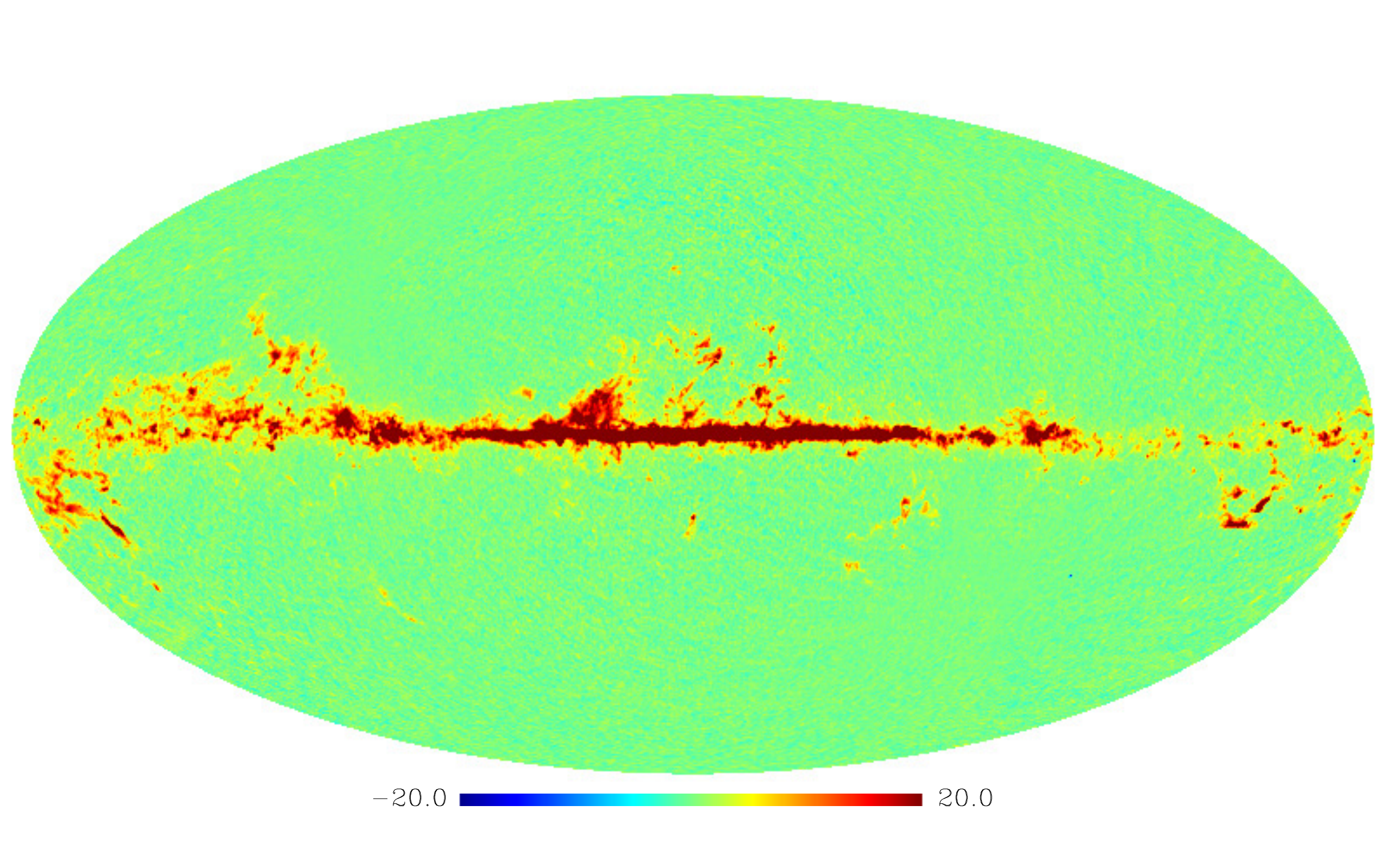}
\caption{MILCA map of $^{12}$CO J=1-0 rotational line at 115 GHz reconstructed from {\it Planck} detectors at 100 GHz at 30' FWHM.}
\label{m12CO}
\end{center}
\end{figure}

\begin{figure}[!th]
\begin{center}
\includegraphics[angle=90,scale=0.35]{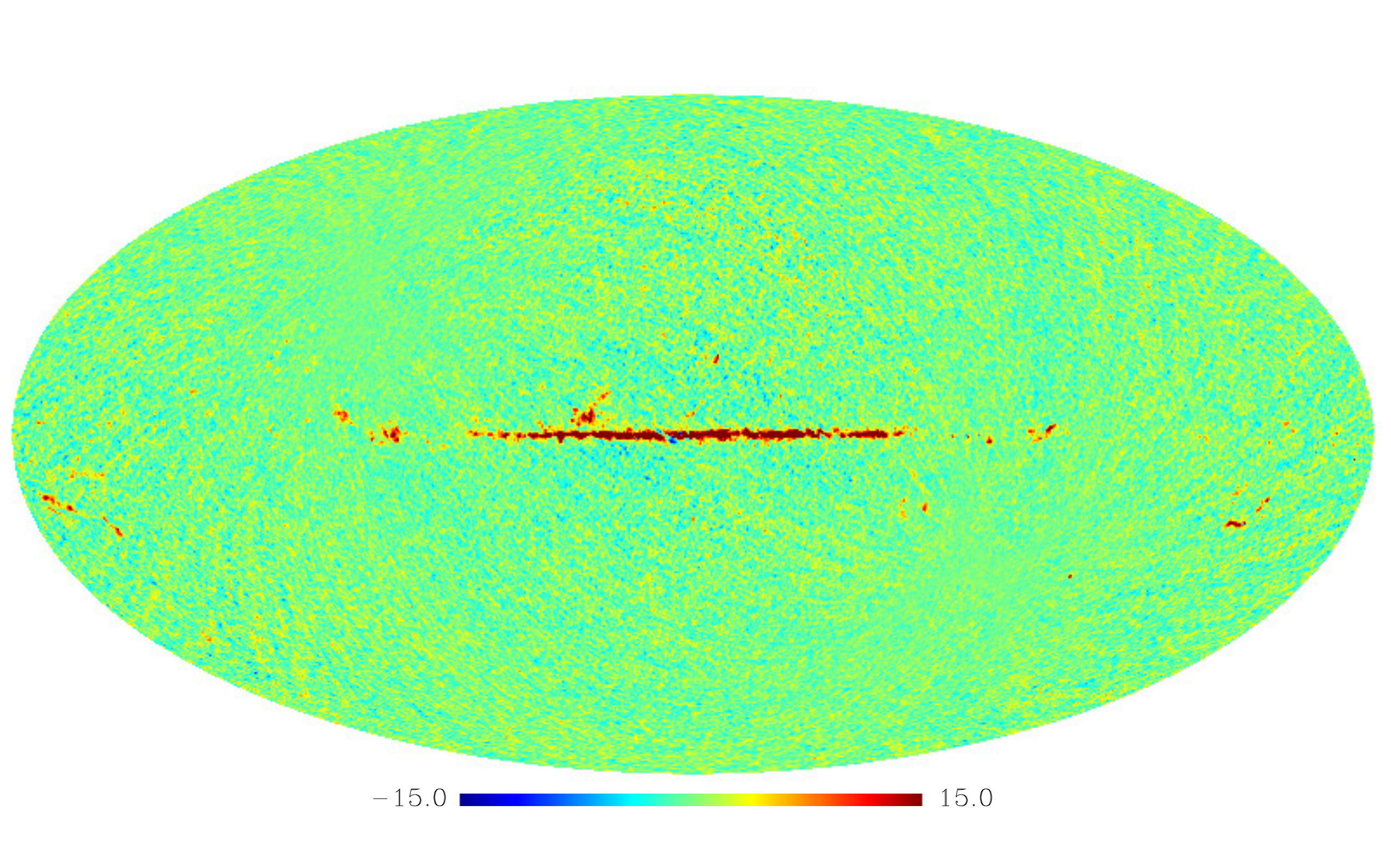}
\caption{MILCA map of $^{13}$CO J=1-0 rotational line at 110 GHz reconstructed from {\it Planck} detectors at 100 GHz at 30' FWHM.}
\label{m13CO}
\end{center}
\end{figure}

\begin{figure*}[!th]
\begin{center}
\includegraphics[angle=0,scale=0.2]{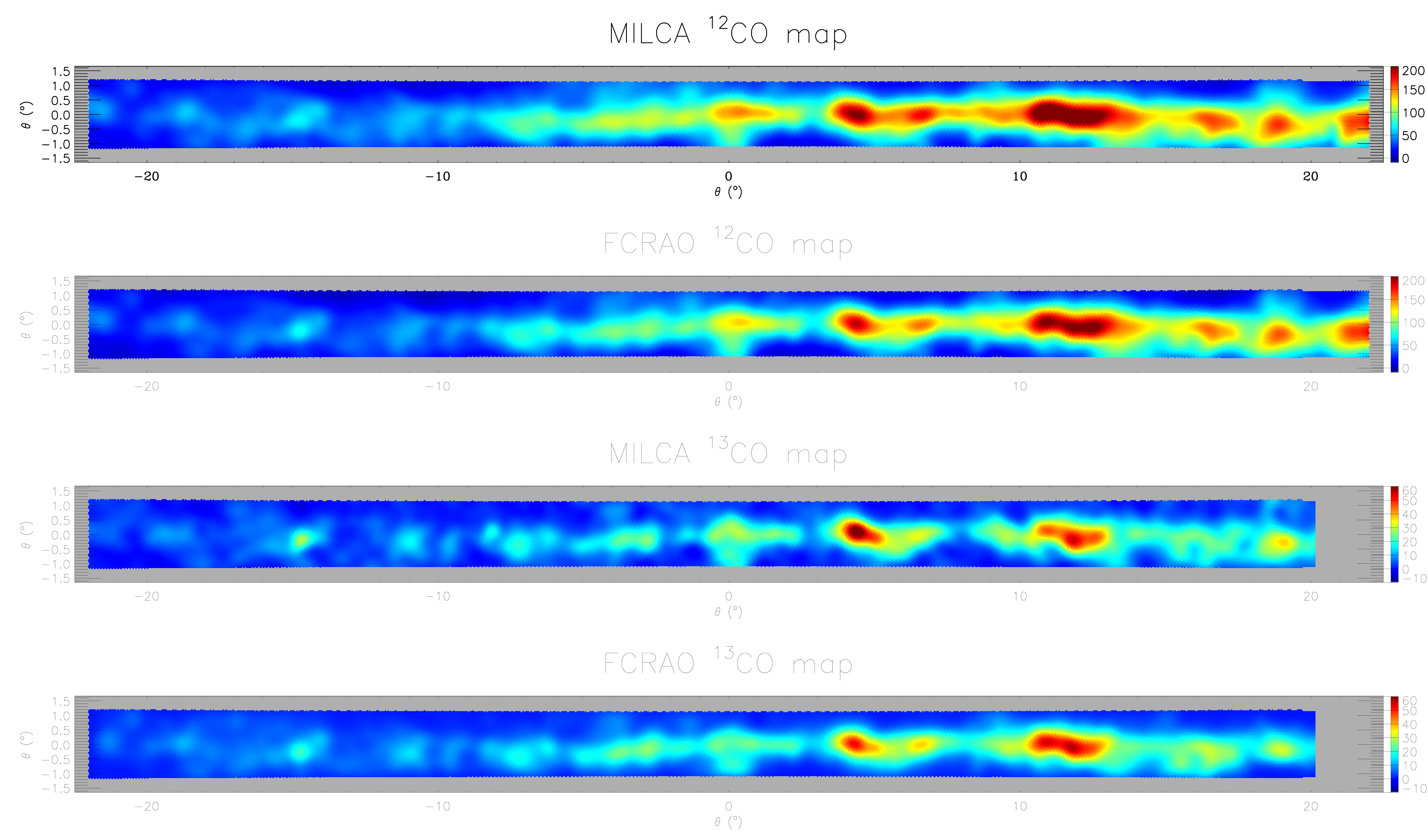}
\caption{From top to bottom, MILCA, FCRAO $^{12}$CO maps, MILCA ,and FCRAO $^{13}$CO maps in K$_{\rm RJ}$.km/s units for the Galactic Ring Survey field \citep{jac06} at 30' FWHM angular resolution.}
\label{grs1213}
\end{center}
\end{figure*}

\subsection{$^{12}$CO and $^{13}$CO response in {\it Planck} maps}
\label{coeffs}
To evaluate the level of both $^{12}$CO and $^{13}$CO in the {\it Planck} type-1 CO\footnote{This map has been obtained from a linear combination of {\it Planck} 100 GHz channel detectors using the MILCA method \citep{hur13}} and \emph{detset}-difference maps, we used \citet{dam01} and BU FCRAO GRS data as tracers of $^{12}$CO and $^{13}$CO.
We describe both the type-1 CO map, $T_{\rm CO}$, and the \emph{detset}-difference map, $T_{\rm \Delta DS} = T_{\rm DS1} - T_{\rm DS2}$, assuming a parametrization of the form,
\begin{align}
T_i = F^{(i)}_{12} \, {\rm Dame} + F^{(i)}_{13} \, {\rm GRS},
\end{align}
and we derive, trough a linear-fit, $F^{\rm CO}_{12} = 0.97 \pm 0.01$ and $F^{\rm CO}_{13} = 0.47 \pm 0.01$ for the type-1 CO map.
For the detset-difference map, we obtain $F^{\Delta \rm DS}_{12} = (1.01 \pm 0.01) \times 10^{-6}$ K$_{\rm RJ}$.km/s/K$_{\rm CMB}$ and $F^{\Delta \rm DS}_{13} = (1.88 \pm 0.01) \times 10^{-6}$ K$_{\rm RJ}$.km/s/K$_{\rm CMB}$

\begin{figure*}[!th]
\begin{center}
\includegraphics[angle=0,scale=0.2]{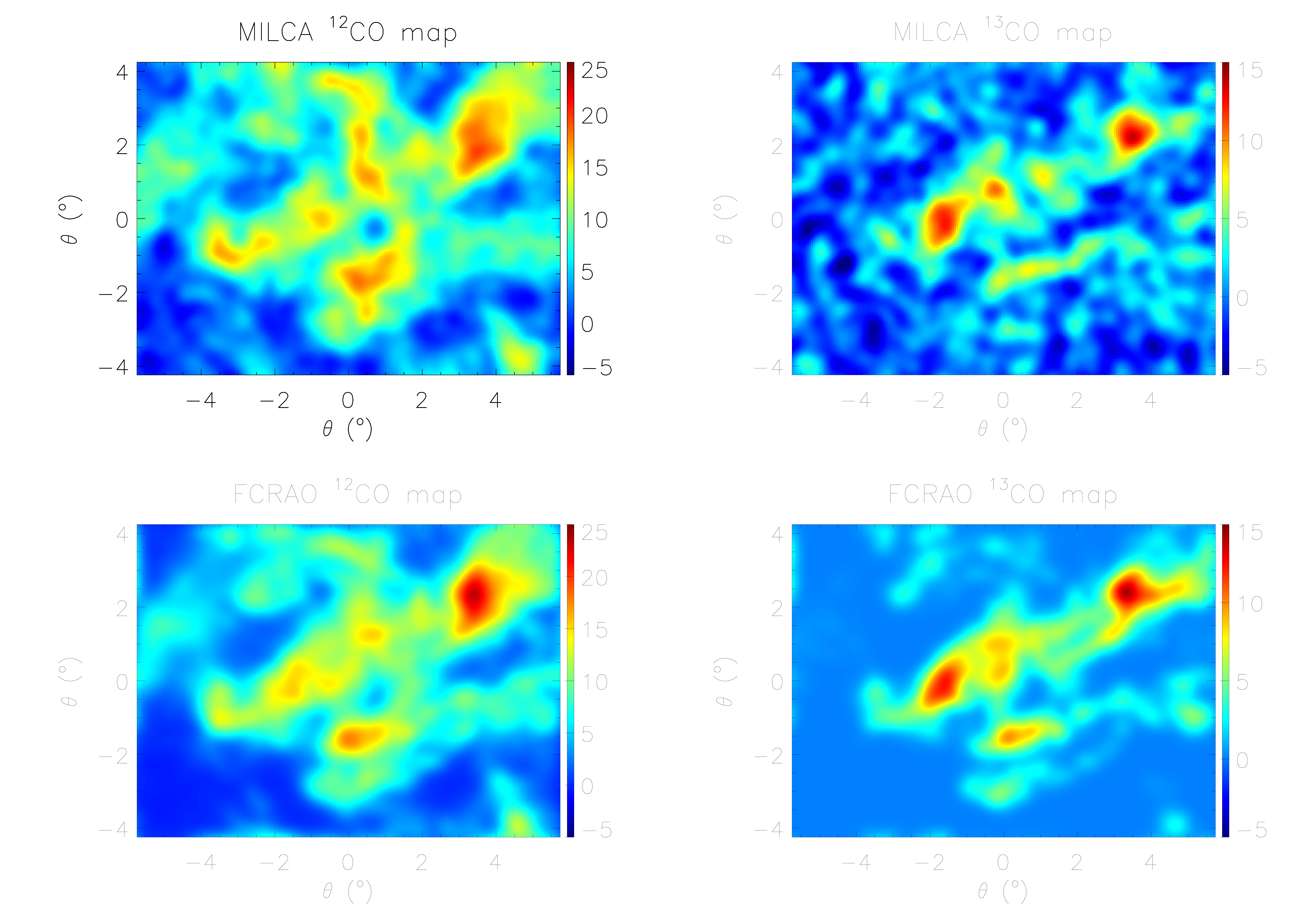}
\caption{From top to bottom, MILCA, FCRAO $^{12}$CO maps, MILCA ,and FCRAO $^{13}$CO maps in K$_{\rm RJ}$.km/s units for the Taurus field \citep{nar08} at 30' FWHM angular resolution.}
\label{taurus1213}
\end{center}
\end{figure*}

\begin{table*}
\center
\caption{Correlation coefficients, uncorrected for instrumental noise covariance, between MILCA CO maps, ancillary observations of CO lines, and thermal dust emission as traced by the {\it Planck} 857 GHz channel map.}
\begin{tabular}{|c|c|cccc|c|}
\hline
\multicolumn{2}{|c|}{$\rho$} & \multicolumn{5}{c|}{Tracers}\\
  \hline
  Region & Maps & MILCA $^{12}$CO  & MILCA $^{13}$CO & FCRAO $^{12}$CO  & FCRAO $^{13}$CO & {\it Planck} 857 Ghz \\ \hline
  \multirow{4}{*}{Galactic Ring Survey} & MILCA $^{12}$CO  & 1.000 & 0.898 & 0.996 & 0.944 & 0.956 \\ 
    & MILCA $^{13}$CO  &0.898&1.000& 0.929 & 0.939 & 0.933 \\ 
    & FCRAO $^{12}$CO  &0.996&0.929& 1.000 & 0.941 & 0.951 \\ 
    & FCRAO $^{13}$CO  &0.944&0.939& 0.941 & 1.000 & 0.946 \\ \hline
  \multirow{4}{*}{Ophiuchus} & MILCA $^{12}$CO &1.000& 0.610 & 0.940 & 0.867 & 0.841 \\ 
    & MILCA $^{13}$CO  & 0.610 &1.000& 0.791 & 0.864 & 0.690 \\ 
    & FCRAO $^{12}$CO  &0.940&0.791& 1.000 & 0.959 & 0.893 \\ 
    & FCRAO $^{13}$CO  &0.867&0.864& 0.959 & 1.000 & 0.888 \\ \hline
  \multirow{4}{*}{Perseus} & MILCA $^{12}$CO &1.000& 0.967 & 0.961 & 0.883 & 0.543 \\ 
    & MILCA $^{13}$CO & 0.967 &1.000& 0.749 & 0.836 & 0.564\\ 
    & FCRAO $^{12}$CO  &0.961&0.749& 1.000 & 0.941 & 0.596 \\ 
    & FCRAO $^{13}$CO  &0.883&0.836& 0.941 & 1.000 & 0.763 \\ \hline
  \multirow{4}{*}{Taurus} & MILCA $^{12}$CO &1.000& 0.453 & 0.939 & 0.719 & 0.691 \\ 
    & MILCA $^{13}$CO & 0.453 &1.000& 0.672 & 0.812 & 0.687 \\  
    & FCRAO $^{12}$CO  &0.939&0.672& 1.000 & 0.853 & 0.758 \\ 
    & FCRAO $^{13}$CO  &0.719&0.812& 0.853 & 1.000 & 0.840 \\
 \hline
\end{tabular}
\label{cormap}
\end{table*}

\subsection{Component separation}

\begin{figure*}[!th]
\begin{center}
\includegraphics[angle=0,scale=0.18]{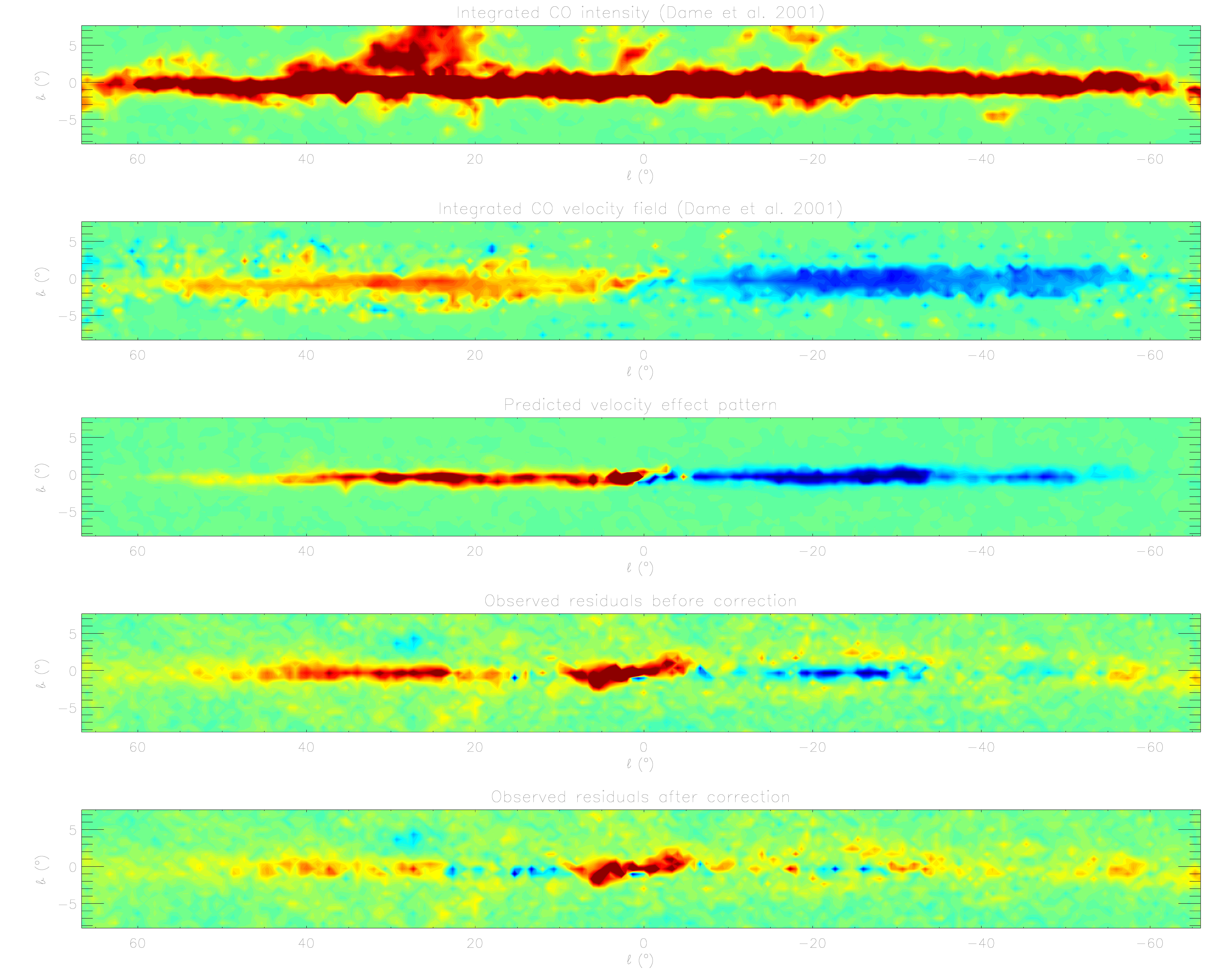}
\caption{From top to bottom : Integrated intensity CO J=1-0 map from \citet{dam01}, Integrated velocity field map from \citet{dam01}, predicted velocity effect pattern, observed residuals between MILCA $^{12}$CO and \citet{dam01} $^{12}$CO maps before and after correction.}
\label{veleff}
\end{center}
\end{figure*}

Using the CO spectral responses of our two CO maps (Type-1 CO MILCA map and \emph{detset}-difference map), we performed a component separation to isolate $^{12}$CO and $^{13}$CO.
In the present situation the component separation is unique and given by the inversion of the mixing matrix derived in Sect.~\ref{coeffs}.
We obtain,
\begin{align}
\label{coweights}
W_{^{12}{\rm CO}} &= 1.390 \times T_{\rm CO} - 3.462 \times 10^5 \times (T_{\rm DS1} - T_{\rm DS2}), \nonumber \\
W_{^{13}{\rm CO}} &= -0.747 \times T_{\rm CO} + 7.185 \times 10^5 \times (T_{\rm DS1} - T_{\rm DS2}).
\end{align}
In the following, we refers to this two maps as MILCA-$^{12}$CO and MILCA-$^{13}$CO maps.
They are presented on Figures~\ref{m12CO}~and~\ref{m13CO}. The two maps are shown at a 30' FWHM angular resolution to enhance the signal to noise ratio of molecular clouds, especially for the $^{13}$CO map. 
We note that the map can be obtained at a resolution of 9.88' which is the native the angular resolution of the {\it Planck} 100 GHz detectors.
Considering that these maps have been derived from a single-channel component separation approach, they do not present a significant leakage from other astrophysical components \citep{PlanckCO}.
We observe that the $^{13}CO$ emission presents a significantly different spatial distribution than the $^{12}$CO emission. The $^{13}$CO is, as expected, mainly observed in regions showing a high $^{12}$CO emission.

\section{Comparison with ancillary data}
\label{seccomp}
In this section, we compare our MILCA $^{12}$CO and $^{13}$CO maps with well known molecular clouds that have already been observed in both $^{12}$CO and $^{13}$CO. We also compare our maps with the thermal dust emission at 857 GHz.
Table.~\ref{cormap} summarize the correlation coefficients of the MILCA $^{12}$CO and $^{13}$CO with reference CO surveys including the BU FCRAO GRS \citep{jac06}, Ophiuchus, Perseus \citep{rid06}, and Taurus \citep{nar08} fields.\\
We observe that the MILCA-$^{12}$CO and the MILCA-$^{13}$CO present a higher level of correlation with ancillary $^{12}$CO and $^{13}$CO data respectively than with thermal dust emission as traced by the {\it Planck} 857 GHz channel.
By construction, these correlations are expected for the BU FCRAO GRS data that has been used to calibrate the CO transmission within the {\it Planck} used in the present analysis.
In Fig.~\ref{grs1213} we present the comparison of the MILCA and BU FCRAO GRS $^{12}$CO and $^{13}$CO maps to illustrate the difference of distribution between $^{12}$CO and $^{13}$CO emission.\\
The other molecular clouds are providing an independent validation that the MILCA $^{12}$CO and $^{13}$CO maps are providing a good separation of the two CO isotopes rotational lines.
In particular, the Taurus molecular cloud shows a high level of correlation for the $^{13}$CO, $\rho = 0.812$, whereas the correlation between the MILCA $^{13}$CO and the FCRAO $^{12}CO$ map is small $\rho = 0.672$. 
The Taurus field is presented on Fig.~\ref{taurus1213}, it illustrates the significantly different morphologies for the $^{12}$CO and $^{13}$CO spatial distributions.\\
The FCRAO $^{12}$CO and $^{13}$CO Taurus maps have a correlation coefficient of $\rho = 0.853$, whereas MILCA $^{12}$CO and $^{13}$CO Taurus maps have a correlation coefficient of $\rho = 0.453$ if not corrected for the instrumental noise covariance and a correlation factor of $\rho = 0.886$\footnote{This value is most likely slightly over-estimated considering that the noise covariance matrix is itself slightly over-estimated due to astrophysical signal residuals in half-mission difference maps. These residuals are produced by small mismatch of line-of-sight within a pixel for the two halves of the mission, and is therefore proportional to the sky intensity gradient.} when correcting for instrumental noise covariance (see Sect.~\ref{seccar} for a description of the instrumental noise estimation), which is consistent with the value found between the $^{12}$CO and $^{13}$CO Taurus maps.\\
It worth noting that MILCA $^{13}$CO and the FCRAO $^{13}CO$ maps are showing lower correlation factor than their $^{12}$CO counterpart due to the significantly lower signal-to-noise ratio of the MILCA-$^{13}$CO map.

\section{Characterization of the maps}
\label{seccar}
\subsection{Uncertainties}

In order to estimate uncertainties, we used the so-called half-mission maps to build an astrophysical signal-free map for each data-map,
\begin{align}
N_{i} =  \frac{1}{2} \left(T^{\rm FIRST}_{i} - T^{\rm LAST}_{i}\right),
\end{align}
where, $T^{\rm FIRST}_{i}$ and $T^{\rm LAST}_{i}$ are the map constructed from half-mission subsets of the {\it Planck} data.\\
These maps provides a fair estimation of the noise in {\it Planck} full sky maps.
Then, we propagate our linear combinations (see Eq.~\ref{coweights}) of {\it Planck} data maps, trough the weights, $w_{i,X}$, to these signal-free maps. We derive CO noise maps,
\begin{align}
N_{^X{\rm CO}} = \sum_i w_{i,X} N_i.
\end{align}

We note that, due to the {\it Planck} scanning strategy, the noise is highly inhomogeneous over the sky. 
To account for this inhomogeneities, we consider the number of observation per pixels $H_{i}$ and we build a variance map for each data map
\begin{align}
V^{N}_{i} = \frac{< N^2_{i} H_{i} >}{H_{i}}
\end{align}
Finally, we propagate these variance maps through our linear combination to deduce CO variance maps,
 \begin{align}
V^{N}_{^X{\rm CO}} = \sum_i w^2_{i,X} V^{N}_i.
\end{align}
At 10' FWHM angular resolution, with a pixel size of 1.7', we obtain an average white noise with a mean standard deviation of 41.2 K$_{\rm RJ}$.km/s for the $^{12}{\rm CO}$ and 70.2 K$_{\rm RJ}$.km/s for the $^{13}{\rm CO}$ map.

\subsection{CO isotopes residual mixing}

The CO transmission in the type-1 CO map and in the \emph{detset}-difference map have been estimated from the data and are therefore affected by uncertainties. 
Uncertainties over this transmission coefficients propagate to the weights of the linear combination used to build the MILCA-$^{12}$CO and the MILCA-$^{13}$CO maps. 
As a consequence, it is expected that the two maps are still presenting some mixing induced by uncertainties over the transmission coefficients.
We performed 1000 Monte-Carlo simulations to estimate the expected level of mixing between the $^{12}$CO and $^{13}$CO components. We derived a level of mixing of $\sim$1\%. Considering that the $^{12}$CO emission has a higher intensity than the $^{13}$CO emission (with a typical line ratio of $\sim 0.2$), the $^{12}$CO is affected by $^{13}$CO contamination at a $\sim$0.2\% level, whereas $^{13}$CO is affected by $^{12}$CO emission at a $\sim$5\% level.
This low-level of contamination is consistent with the $^{12}$CO and $^{13}$CO correlation factor in the Taurus field once corrected for instrumental noise covariance, which is consistent with the correlation level of the FCRAO Taurus maps. Leakage between the two component would produce a significantly lower value for this correlation\footnote{This lower level of correlation would be induced by the anti-correlation of the uncertainty over the transmission coefficients for $^{12}$CO and $^{13}$CO.}

\subsection{Velocity effect}

The doppler effect on a CO line induces a modification of the observation frequency, and thus a modification of the effective response induced by the gradient of the CO map effective bandpass.
The observed CO integrated intensity in the MILCA maps can be express as,
\begin{align}
W^{\rm obs}_{\rm CO} = \frac{\left(1+v_{\rm CO}/g \right)}{A_0} W^{\rm real}_{\rm CO},
\end{align}
where $g$ is the inverse of the gradient of the bandpass, $A_0$ denotes the mismatch on absolute calibration of the map due to non-zero velocity on the line of sight for the sky region (BU FCRAO GRS covered sky) used to calibrate the map, and $v_{\rm CO}$ is the mean velocity of the CO on the line of sight.
We construct a map of the CO velocity field from \citet{dam01} cube data,
\begin{align}
W^{\rm obs}_{\rm CO} = \int I_{\rm CO}(u)\, {\rm d}u, \nonumber \\
v_{\rm CO} = \frac{\int u\, I_{\rm CO}(u) \, {\rm d}u}{W^{\rm obs}_{\rm CO}},
\end{align}
where $I_{\rm CO}(u)$ is the CO intensity at a velocity $u$ on the line of sight.\\

In Fig.~\ref{veleff}, we present in the two upper panels the $W^{\rm obs}_{\rm CO}$ and $v_{\rm CO}$ maps computed from \citet{dam01} data. In the third panel from the top, we present the predicted pattern of the velocity effect, $R \propto W^{\rm obs}_{\rm CO}\,v_{\rm CO}$, with $R$ the residual emission in the difference map between \citet{dam01} and MILCA $^{12}$CO map. This residual is shown on the antepenultimate panel of Fig.~\ref{veleff}.\\
Finally, we fit the bandpass gradient, $g = (616 \pm 1)$ km/s and the calibration mismatch $A_0 = 1.0 \pm 0.1$. The bottom panel of Fig.~\ref{veleff} presents the residuals after correction of the velocity effect.\\
 
Considering a nearly constant $^{13}$CO(J=1-0)/$^{12}$CO(J=1-0) ratio in the galactic disc, we can estimate the velocity effect on the MILCA $^{13}$CO map. The effect for this map is consistent with zero as no clear feature in the $^{13}$CO(J=1-0)/$^{12}$CO(J=1-0) ratio is correlated to the velocity field.

\subsection{Calibration}

Using our method, we only have access to CO relative transmission. 
Consequently, by construction we have the same calibration than \citet{dam01} and \citet{jac06} for $^{12}$CO and $^{13}$CO respectively.

\begin{figure}[!th]
\begin{center}
\includegraphics[angle=0,scale=0.2]{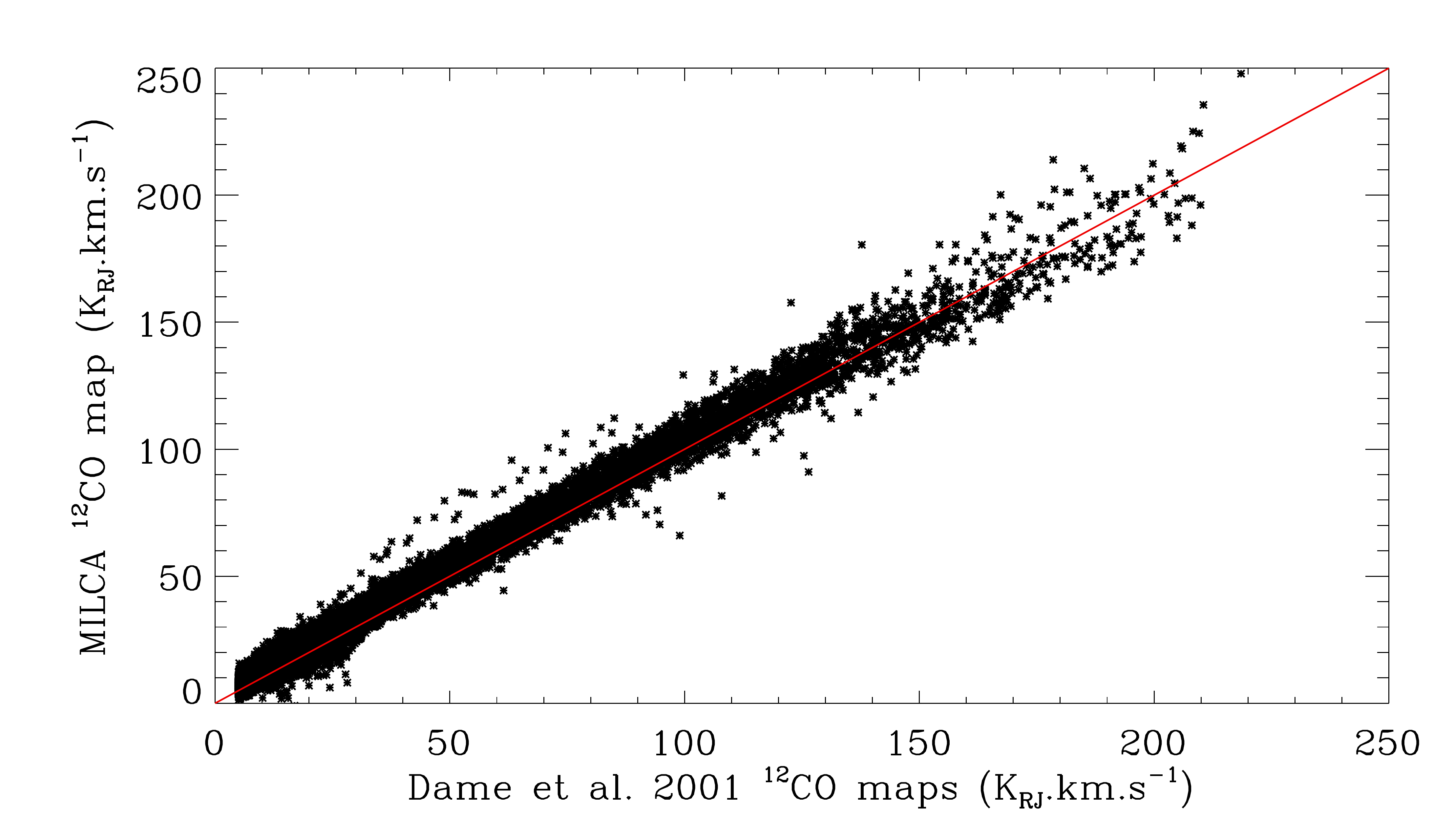}
\caption{Correlation between MILCA $^{12}$CO map after correction from velocity effect and \citet{dam01} map. The red line shows the 1:1 correlation.}
\label{cor12a}
\end{center}
\end{figure}

\begin{figure}[!th]
\begin{center}
\includegraphics[angle=0,scale=0.2]{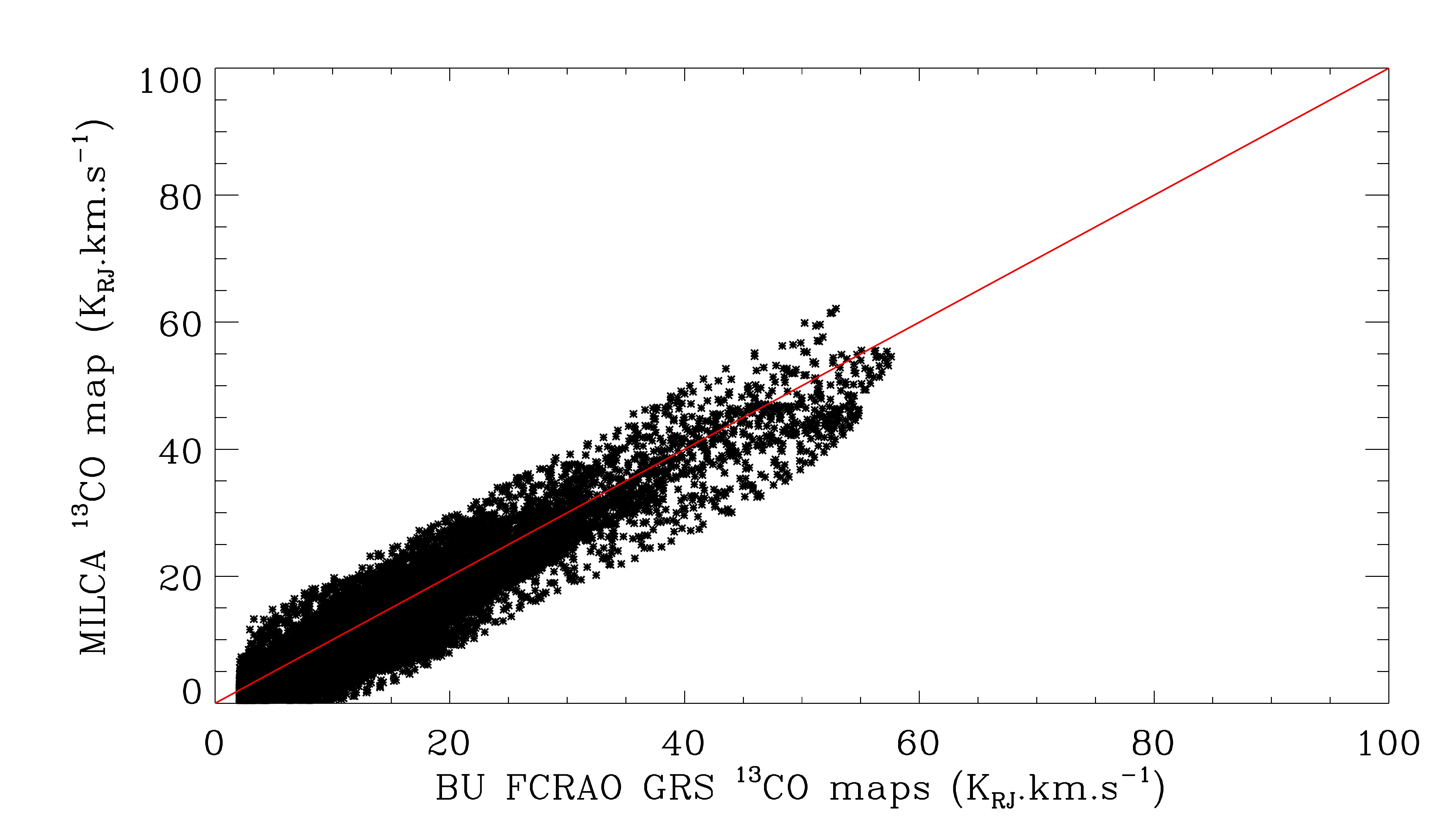}
\caption{Correlation between MILCA $^{13}$CO map and BU FCRAO GRS $^{13}$CO map. The red line shows the 1:1 correlation.}
\label{cor13}
\end{center}
\end{figure}

We compute the correlation between MILCA $^{12}$CO, MILCA $^{13}$CO and \citet{dam01} and BU FCRAO GRS maps.\\
In Fig.~\ref{cor12a}, we present the correlation between MILCA $^{12}$CO and \citet{dam01} map after correction of the velocity effect. We observe a 1:1 correlation between the two datasets.
In Fig.~\ref{cor13}, we also observe a 1:1 correlation between MILCA $^{13}$CO and BU FCRAO GRS map.

\begin{figure}[!th]
\begin{center}
\includegraphics[angle=0,scale=0.2]{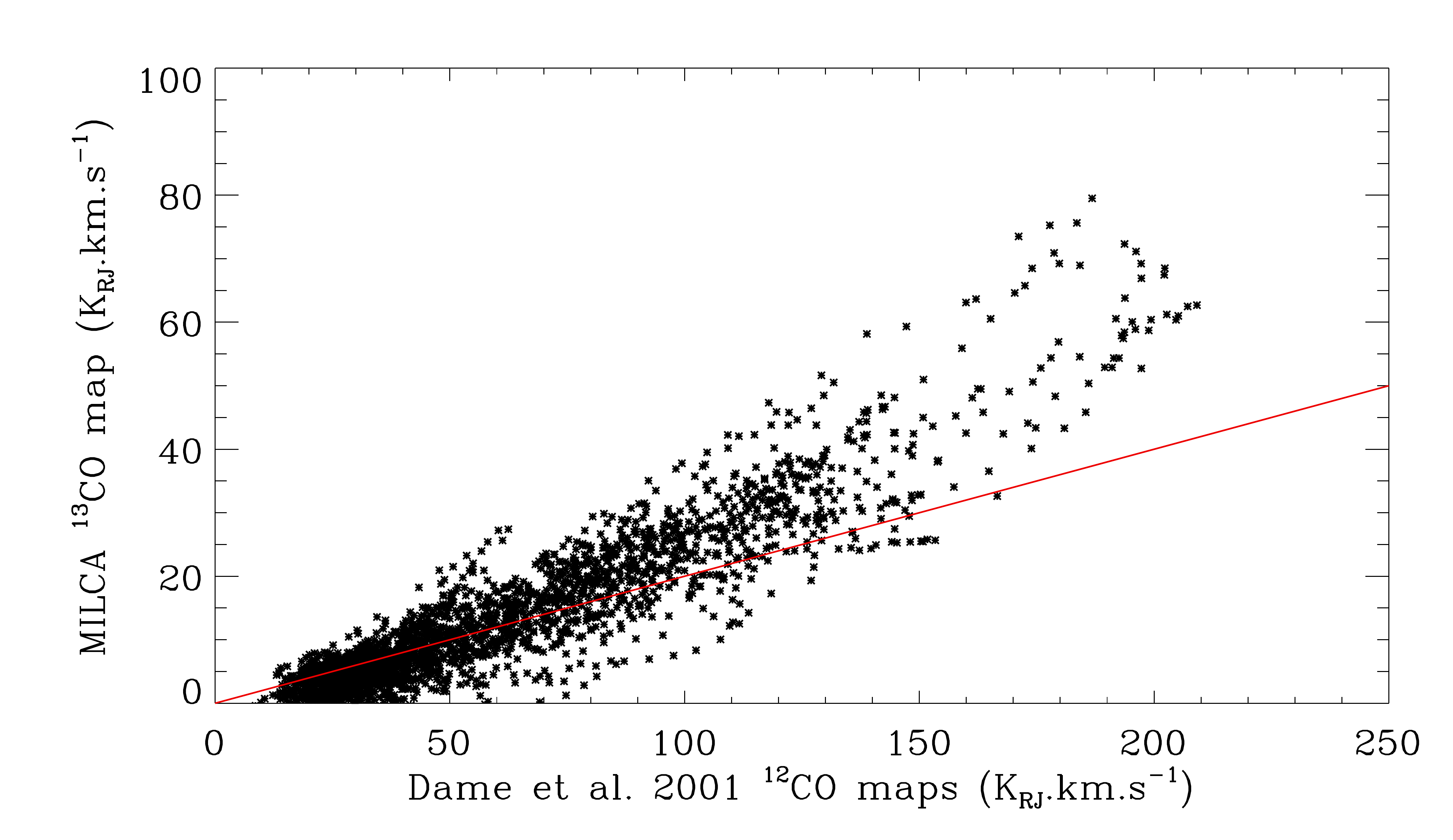}
\caption{Correlation between MILCA $^{13}$CO map and and \citet{dam01} map. The red line shows a 5:1 slope.}
\label{cor1213}
\end{center}
\end{figure}

For comparison, we also correlated the MILCA $^{13}$CO with the \citet{dam01} map. We show this correlation in Fig.~\ref{cor1213}.\\
We observe, that the relation is not linear. As expected, the faint $^{12}$CO regions present a low $^{13}$CO(J=1-0)/$^{12}$CO(J=1-0) ratio, whereas high intensity regions present a higher isotope ratio.

\section{Improving the maps in the future}
\label{secdis}
We note that once/if released individual detector maps (or even Time Ordered Information, TOI) at 100 GHz will allow to build CO maps with a lower noise level (An improvement of the order of 20-40\% for the noise level compared to the map presented in this paper might be expected and will be determined by the exact distribution of CO transmission coefficients in the 8 {\it Planck} detectors at 100 GHz).\\
An optimal $^{12}$CO and $^{13}$CO separation should be performed at the map-making level using a nine components model: I, Q, U stokes parameters for the 100 GHz sky, the $^{12}$CO and the $^{13}$CO maps.
It will however requires the derivation of the $^{12}$CO and $^{13}$CO transmission coefficients for all detectors (8 at 100 GHz) and the gradient of the transmission curves at 110 and 115 GHz for all detectors (to account for transmission variation due the satellite motion with respect to the molecular clouds). The determination of such CO coefficients (16 in totals) will be challenging as it will require to use CO templates at least in intensity and maybe in polarization to avoid biases on those coefficients induced by polarized CO emission (that could be handled by performing the coefficient measurement in low polarized fraction regions).\\
It is important to precise that for the 100 GHz the possibility of extracting polarized CO emission will be limited by the number of detectors (8) and will require to use the difference of the scanning strategy between {\it Planck} surveys 1/3 and 2/4 to be able to have a sufficient number of measurements to extract the 9 components. Otherwise, polarized CO emission would be significantly biased by residual leakage from $^{12}$CO and $^{13}$CO intensity. This also imply that for regions of the sky not showing significant difference for the scanning strategy between surveys 1/3 and 2/4 the $^{12}$CO and $^{13}$CO polarization will present a very high-level of noise reconstruction.\\

We also note that the same procedure could be used at 217 and 353 GHz to extract $^{12}$CO and $^{13}$CO for the J=2-1 and J=3-2 rotational lines. However, it will also require templates to extract transmission coefficients in the individual detector maps for $^{12}$CO and $^{13}$CO separately. The construction of such map will be let for future work.\\
Finally, it is important to note that the 217 GHz channel is well suited for polarized CO extraction as it possesses 12 detectors (4 unpolarized + 8 polarized) for 9 components (3 intensity + 6 polarization) to extract.\\
The case of the 353 GHz will be more challenging as thermal dust bandpass mismatch is not negligible, and thus a 12 components (4 intensity + 8 polarization) model is required to describe the 12 detectors (4 unpolarized + 8 polarized), which may lead to very noisy reconstructed maps.

\section{Conclusion}
\label{seccon}
We have presented and applied a method to construct separated $^{12}$CO and $^{13}$CO maps from public {\it Planck} data combining the Type-1 CO MILCA map with detector set maps at 100 GHz.
This work present the first full sky cartography for the $^{13}$CO(J=1-0) line.\\
We have carefully characterized statistical uncertainties and systematic effect within the map.
We also compared our MILCA CO maps with ancillary data. This comparison illustrate that we have effectively separated $^{12}$CO and $^{13}$CO.
In the future improvements on these maps are expected when all {\it Planck} data will be public

\section*{Acknowledgment}
\thanks{
\footnotesize G.H. acknowledge support from Spanish Ministerio de Econom\'ia and Competitividad (MINECO) through grant number AYA2015-66211-C2-2.
We acknowledge the use of HEALPix \citep{gor05}. This publication makes use of molecular line data from the Boston University-FCRAO Galactic Ring Survey (GRS). The GRS is a joint project of Boston University and Five College Radio Astronomy Observatory, funded by the National Science Foundation under grants AST-9800334, AST-0098562, \& AST-0100793. \\ 
    }

\bibliographystyle{aa}
\bibliography{CO1213}

\end{document}